\documentclass[twocolumn,preprintnumbers,amsmath,amssymb]{revtex4-1}
\usepackage{epsfig}
\usepackage{amsmath}
\usepackage{MnSymbol}
\usepackage{siunitx}
\usepackage{soul}
\usepackage{array}
\usepackage{multirow}
\usepackage{psfrag}
\usepackage[font=footnotesize,format=plain,labelsep=period,justification=raggedright]{caption}
\usepackage{subcaption}
\usepackage{xcolor}
\usepackage{graphicx}

\begin{document}
\title{Factors Controlling the Pinning Force of Liquid Droplets on Liquid Infused Surfaces}
\author{Muhammad Subkhi Sadullah, Jack R. Panter, and Halim Kusumaatmaja$^{\ast}$}
\affiliation{Department of Physics, Durham University, Durham, DH1 3LE, UK.}
\affiliation{$^{\ast}$Email: halim.kusumaatmaja@durham.ac.uk}

\begin{abstract}
	Liquid infused surfaces with partially wetting lubricants have recently been exploited for numerous intriguing applications, such as for droplet manipulation, droplet collection and spontaneous motion. When partially wetting lubricants are used, the pinning force is a key factor that can strongly affect droplet mobility. Here, we derive an analytical prediction for contact angle hysteresis {in the limit where the meniscus size is much smaller than the droplet}, and numerically study how it is controlled by the solid fraction, the lubricant wetting angles, and the various fluid surface tensions. We further relate the contact angle hysteresis and the pinning force experienced by a droplet on a liquid infused surface, and our predictions for the critical sliding angles are consistent with existing experimental observations. Finally, we discuss why a droplet on a liquid infused surface with partially wetting lubricants typically experiences stronger pinning compared to a droplet on a classical superhydrophobic surface.
\end{abstract}

\maketitle
\renewcommand{\arraystretch}{1.5}

\section{Introduction}
Since their inception \cite{Quere2005,Wong2011,Smith2013}, liquid infused surfaces (LIS) have been prized for their anti-adhesive nature, which results in properties such as the high mobility of liquid droplets and anti-fouling. These properties are highly desirable in a broad range of applications, from marine and medical coatings \cite{Sotiri2016,Ware2018}, to non-stick packaging \cite{Brown2017}, and digital microfluidics \cite{Geng2019}.

LIS are formed by impregnating a rough, porous or textured surface with a lubricating liquid, which is immiscible to the mobile liquid phase introduced to LIS. {The lubricant also needs to preferentially wet the solid compared to the mobile liquid phase \cite{Smith2013}.} This lubricant layer imbues LIS with significant advantages over superhydrophobic surfaces, such as pressure stability \cite{Wong2011} and self healing \cite{Wei2014}.

High mobility of a liquid droplet is particularly obtained when the lubricant completely wets the surface texture, as pinning of the droplet on the surface is negated by the intervening lubricant layer. However, the dependency on a fully wetting lubricant often limits the implementation of LIS, both due to the difficulty of finding the suitable lubricant for the desired applications \cite{Smith2013,Galvan2018}, and due to the possibility of lubricant depletion \cite{Howell2015, Liu2016,Sett2017,Peppou-Chapman2018,Kreder2018}.

On the other hand, LIS with partially wetting lubricant have increasingly attracted interest, especially with a number of external stimuli shown to allow reversible change of wetting states from slippery to sticky (see for example the recent review \cite{Lou2020}). {Such surfaces have substantially expanded functionality compared to the purely slippery surfaces,} with the ability to locally change a droplets' mobility leading to the demonstration of fog capture even in high winds \cite{Wang2017}, to introduce bidirectional motion under texture gradients \cite{Sadullah2020}, and recently the unprecedented manipulation of both droplets and colloids \cite{Wang2018}.

It is important therefore to understand pinning from two perspectives: as a problem to be minimised, or as a functional phenomenon to be controlled. However, {to the best of our knowledge}, the quantitative relationship of the pinning force to both the surface roughness and fluid properties has never been systematically studied on LIS.

{In this contribution, we study the pinning force and contact angle hysteresis (CAH) of a droplet on LIS. We begin by developing an analytical model for CAH based on averaged Cassie-Baxter surface properties. For simplicity, we limit our study to the case where the meniscus size is significantly smaller than the droplet size.} We then observe both advancing and receding contact angles using computer simulations, showing that the simulated CAH closely matches the analytical results. We, therefore, are able to accurately quantify the hysteresis based on the surface roughness, and the set of fluid-solid and fluid-fluid surface tensions. Further, we derive the total pinning force and demonstrate that our prediction is consistent with experimental observations.

We find{, similar to droplets on solid surfaces,} that there is a competition between two factors which control the pinning force on LIS, (i) the droplet base perimeter and (ii) the cosine difference between the receding and the advancing angles. Our theory suggests that this competition minimises pinning {when the apparent contact angle approaches $\theta_{app}\rightarrow 0^\circ$ or $\theta_{app}\rightarrow 180^\circ$, but maximises it at an intermediate value of the apparent contact angle, $\theta_{app}\simeq 65.5^\circ$. Since most reported values of the apparent contact angle on LIS are moderate ($\theta_{app} \sim 80-100^\circ$), this means a small but non-zero CAH can lead to a significant critical sliding angle for droplets on LIS.}

\section{Methods}
In this work, we are interested in static wetting configurations, rather than the dynamics of LIS system. The static configurations can be obtained by minimising the total free energy. A typical LIS system consists of three fluid phases (droplet, lubricant and gas phases) and a textured solid. To simulate such systems, we need a free energy model which can describe (i) the existence of the three fluid phases and their respective interfacial tensions, and (ii) fluid-solid interactions, which in turn determines the contact angles of the fluids on the solid surface.

To do this, we employ a diffuse interface approach in which the total free energy is constructed as
\begin{align}
\Psi = \int_V \Psi_{fluid} \, \mathrm{d}V + \int_S \Psi_{surface} \, \mathrm{d}S.
\label{eq:freeenergy1}
\end{align}
The fluid free energy is given by \cite{Semprebon2016a}
\begin{align}
\Psi_{fluid} &= \sum_{m=1}^{3} \frac{\kappa_m}{2} C_m^2 (1 - C_m)^2 +\frac{a^2\kappa_m}{2} (\nabla C_m)^2.
\label{eq:fluid_bulk_interface}
\end{align}
The $C_m$'s are the order parameters which represent the co-existence of three bulk fluid phases in the simulation space. The interfacial tensions between fluid $m$ and $n$ are controlled by the $\kappa_m$ parameters and the interface width $a$ via 
\begin{equation}
\gamma_{mn}=a\left(\kappa_m+\kappa_n\right)/6. 
\end{equation}
For concreteness, we have chosen phases $m =$ 1,2, and 3 to be the droplet (d), gas (g) and lubricant (l) phases respectively. 

The surface free energy is given by \cite{Connington2013}
\begin{align}
\Psi_{surface}&=\sum_{m=1}^{2}-6\gamma_{3m}\cos{\theta_{3m}}\left(\frac{1}{2}C_m|_s^2-\frac{1}{3}C_m|_s^3\right).
\label{eq:cubic_ternary}
\end{align}
Here, $C_m|_s$ is the value of $C_m$ at the surface. We also chose to control the solid-fluid interaction using the parameters $\theta_{31}$ and $\theta_{32}$, corresponding to the lubricant-droplet and lubricant-gas contact angles, $\theta_{ld}$ and $\theta_{lg}$, respectively. For thermodynamic consistency, the third contact angle, i.e. the droplet-gas contact angle, is prescribed once the other two contact angles are determined, following
\begin{equation}
\gamma_{gl} \cos \theta_{gl} + \gamma_{ld} \cos \theta_{ld} + \gamma_{dg} \cos \theta_{dg} =0.
\end{equation}
This is often known as the Good-Girifalco relation \cite{Girifalco1957}.

In this work, the total free energy of the system is minimised using the L-BFGS algorithm \cite{Liu1989}, following the numerical scheme discussed in Refs. \cite{Kusumaatmaja2015, Panter2019}. The L-BFGS algorithm is chosen due to its efficiency for minimisation problems with a large number of degrees of freedom, though in principle other minimisation routines may also be employed.

\section{Results and Discussions}
\subsection{Derivation of Pinning Force and CAH}
The pinning force per unit length for a droplet on a dry (not lubricated) textured solid surface is given by \cite{Quere2008}
\begin{align}
f_{dry}=\gamma_{dg}\Delta\cos\theta,\label{eq:fpinning0}
\end{align}
where $\gamma_{dg}$ is the surface tension of the droplet with the gas phase and $\Delta\cos\theta=\left(\cos{\theta^{R}}-\cos{\theta^{A}}\right)$ is the difference in the cosine of the receding $\theta^{R}$ and the advancing $\theta^{A}$ contact angles for the droplet-gas-solid contact line.

\begin{figure}
	\centering
	\includegraphics[keepaspectratio]{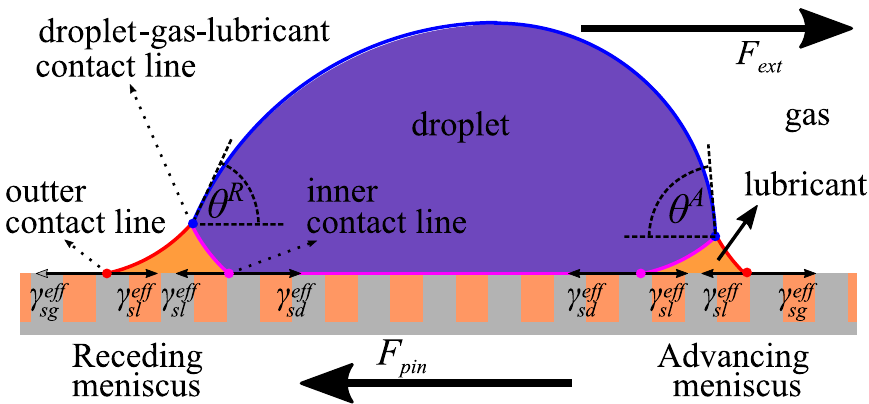}
	\caption{Droplet on LIS under influence of an external force $F_{ext}$. The resisting force due to contact line pinning, $F_{pin}$, is pointing in the opposite direction. We have also shown the surface tension forces acting on the inner and outer contact lines.}
	\label{fig:fig1}
\end{figure}

Compared to other surfaces, the distinguishing feature of LIS is the presence of the lubricant meniscus. As such, the droplet-gas-solid contact line is not present. Instead, we have to consider the compound effect of droplet-gas-lubricant, droplet-lubricant-solid and lubricant-gas-solid contact lines, as illustrated in Fig.~\ref{fig:fig1}. Semprebon \textit{et al.} have derived an expression for the CAH on LIS using geometrical analysis of the meniscus \cite{Semprebon2016b}. Here, we will show that the CAH can also be derived employing a simpler argument based on force balance. 

\begin{figure*}
	\centering
	\includegraphics[width=\linewidth]{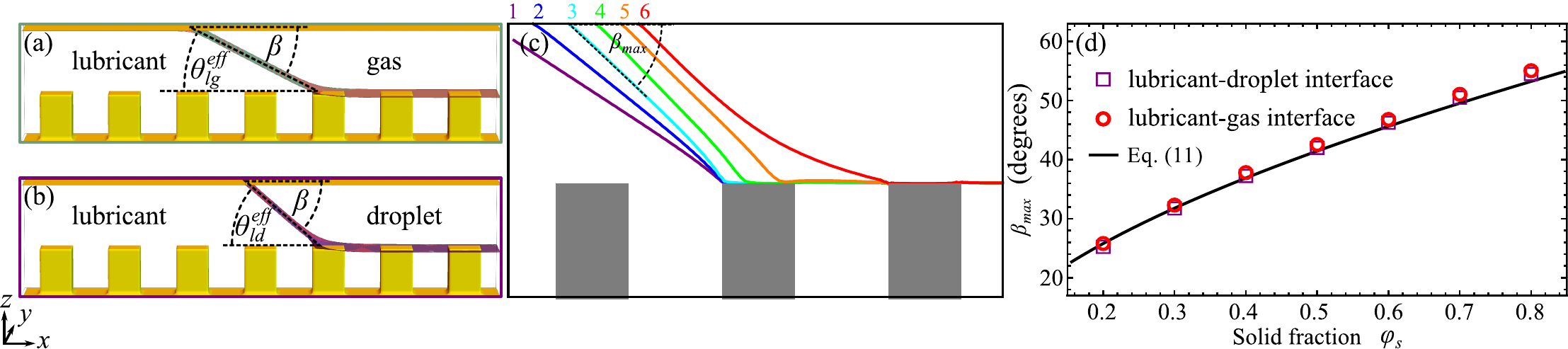}
	\caption{Simulations of the depinning mechanism using quasi 3D setups for (a) the advancing lubricant-gas and (b) the receding lubricant-droplet interfaces. The top contact angle $\beta$ can be tuned to find $\theta_{lg}^{depin}$ and $\theta_{ld}^{depin}$. (c) The typical evolution of the contact line when $\beta$ is increased. For $\beta<\beta_{max}$ (purple and blue lines), the contact line is pinned. At $\beta=\beta_{max}$ (cyan, green, orange and red lines), the contact line slides on top of the post. The measurement of $\beta_{max}$ for different $\varphi_s$ and its comparison with Eq. \eqref{eq:Cassie_Baxter} are presented in panel (d). Here, we have used $\theta_{ld}=\theta_{lg}=60^\circ$.}
	\label{fig:fig2}
\end{figure*}

Let us consider a droplet on LIS under the influence of an external force $F_{ext}$, as shown in Fig. \ref{fig:fig1}.
Here, $\gamma_{sm}^{eff}$ denotes the effective interfacial tension of LIS with the fluid phase $m$, with $m= d, g, l$.
The subscript $s,d,g$ and $l$ are to indicate the solid, droplet, gas and lubricant phases respectively. LIS can be considered as a composite surface where $\varphi_s$ fraction of the surface is solid surface and the remaining $(1-\varphi_s)$ is the lubricant surface \cite{CassieBaxter1944}. The effective interfacial tension of the fluid phase phase $m$ with the composite surface can then be written as
\begin{align}
\gamma_{sm}^{eff}=\varphi_s\gamma_{sm}+(1-\varphi_s)\gamma_{lm}.
\end{align}

The pinning force per unit length for a droplet on LIS $f_{LIS}$ can be calculated from the sum of the effective interfacial tensions of this composite surface at the outer and the inner contact lines, as indicated in Fig.\ref{fig:fig1}. Hence, $f_{LIS}$ is written as
\begin{align}
f_{LIS}=&\left(\left[\gamma_{sg}^{eff}-\gamma_{sl}^{eff}\right]^{R}+\left[\gamma_{sl}^{eff}-\gamma_{sg}^{eff}\right]^{A}\right)_{outer}+\nonumber\\
&\left(\left[\gamma_{sl}^{eff}-\gamma_{sd}^{eff}\right]^{R}+\left[\gamma_{sd}^{eff}-\gamma_{sl}^{eff}\right]^{A}\right)_{inner}.
\label{eq:fpinning1}
\end{align}
The superscripts $A$ and $R$ indicate the advancing and the receding menisci. 

Generally, the terms for the outer and inner contact lines cannot simply be added together since they are to be integrated over different lengths (i.e. the inner and the outer droplet base perimeters). However, in the limit where the meniscus size is small compared to the droplet, the outer and the inner droplet base perimeters can be taken to be approximately the same \cite{Semprebon2016b,Keiser2017}. In this approximation, we can introduce the effective lubricant wetting angles as
\begin{align}
\cos{\theta_{lg}^{eff}}=\frac{\gamma_{sg}^{eff}-\gamma_{sl}^{eff}}{\gamma_{lg}},\quad\cos{\theta_{ld}^{eff}}=\frac{\gamma_{sd}^{eff}-\gamma_{sl}^{eff}}{\gamma_{ld}},
\end{align} 
such that Eq. \eqref{eq:fpinning1} can be written into
\begin{align}
f_{LIS}=\Big(&\left[\gamma_{lg}\cos{\theta_{lg}^{eff}}-\gamma_{ld}\cos{\theta_{ld}^{eff}}\right]^R\nonumber\\-&\left[\gamma_{lg}\cos{\theta_{lg}^{eff}}-\gamma_{ld}\cos{\theta_{ld}^{eff}}\right]^A\Big).
\label{eq:fpinning2}
\end{align}
{The advantage of this expression is that it allows us to easily distinguish the contributions from the advancing and receding menisci. Alternatively, Eq. \eqref{eq:fpinning2} can be further simplified to
	\begin{eqnarray}
	f_{LIS}&=& \gamma_{lg} \left( \cos{\theta_{lg}^{eff,R}} - \cos{\theta_{lg}^{eff,A}}\right) -  
	\gamma_{ld} \left( \cos{\theta_{ld}^{eff,R}} - \cos{\theta_{ld}^{eff,A}} \right)\nonumber \\
	&=& \gamma_{lg} \Delta \cos{\theta_{lg}^{eff}} - \gamma_{ld} \Delta \cos{\theta_{ld}^{eff}}.
	\label{eq:fpinning3}
	\end{eqnarray}
}

\subsubsection{Depinning Mechanisms}
Numerous simulation studies have been conducted to investigate the effective contact angles when a droplet is about to move on a dry textured surface \cite{Mognetti2010,Panter2019}. In such cases, we typically consider two contact line depinning mechanisms, corresponding to the advancing and receding contact lines of the droplet-gas interfaces. In contrast, for LIS, we must consider how both the lubricant-droplet and lubricant-gas interfaces advance and recede.

For an advancing contact line on a dry textured surface, the front part of the droplet typically advances by {\it bridging} to the front subsequent post. For LIS, such a contact line depinning mechanism is also observed for the lubricant-droplet interface at the advancing meniscus as well as the lubricant-gas interface at the receding meniscus \cite{Schellenberger2016}. Therefore, the effective contact angle for both interfaces are zero when they depin, $[\theta_{lg}^{eff}]^R=0$ and $[\theta_{ld}^{eff}]^A=0$. 

There are various mechanisms for the receding contact line to depin from the post, which depend on the post geometry \cite{Panter2019}. For LIS, this is relevant for understanding the lubricant-droplet interface at the receding meniscus and the lubricant-gas interface at the advancing meniscus. Here we will focus on a square array of {rectangular} posts, and we can use our numerical approach to determine the relevant depinning mechanism. To do this, we start by simplifying the system studied and isolate the advancing lubricant-gas and the receding lubricant-droplet interfaces, as shown in Fig. \ref{fig:fig2}(a-b). To reduce the computational cost, we concentrate our numerical study in the region close to the contact line. {We have used a quasi 3D setup where only a single row of posts are explicitly simulated at the bottom surface, a smooth wall is used for the top surface, and periodic boundary condition is applied in the direction perpendicular to the row of posts. Two fluid phases are then introduced in each half of the simulation domain, and the two phases have equal pressure such that their interface is flat. The top contact angle $\beta$ can be controlled to measure the depinning angles. This is performed by varying $\beta$} and recording its critical angle, $\beta_{max}$, for the stability of the corresponding interfaces. Simple geometry then dictates that $\beta_{max}$ is the critical depinning angle for $[\theta^{eff}_{lg}]^A$ and $[\theta^{eff}_{ld}]^R$.

The typical development of a receding interface is shown in Fig. \ref{fig:fig2}(c) upon varying $\beta$. The interface is initially stable and pinned at the corner of the {rectangular} post (purple line). Increasing $\beta$ deforms the interface (blue line) until we eventually reach $\beta_{max}$ (cyan line). Here the interface detaches from the corner and the contact line slides on top of the post (see green, orange and red lines), while maintaining a constant contact angle at the top plate. For this depinning mechanism, the critical angle is given by $\beta_{max} = \theta_{lm}^{CB}$, where $\theta_{lm}^{CB}$ is the Cassie-Baxter contact angle  {\cite{CassieBaxter1944,Priest2007}} of the composite surface:
\begin{align}
\cos \beta_{max}= \cos\theta_{lm}^{CB}=\varphi_s\cos\theta_{lm}+(1-\varphi_s),
\label{eq:Cassie_Baxter}
\end{align}
and $m= d, g$. {For the quasi 3D setup, where the contact lines of the lubricant-gas and the lubricant-droplet interfaces are along the $y$-direction (see Fig. \ref{fig:fig2}), the relevant solid fraction is taken to be the line average, instead of the surface average \cite{Panter2019}. Hence, we define the solid fraction as $\varphi_s=W_y/L_y$, where $W_y$ and $L_y$ are the width and periodicity of the posts in the $y$-direction.} As an illustrative example, Fig. \ref{fig:fig2}(d) shows the measured $\beta_{max}$ for different $\varphi_s$ and Young's angles $\theta_{ld}=\theta_{lg}=60^\circ$. We consistently find this depinning mechanism to be at play for the surface textures considered in this work (square arrays of {rectangular} posts). Similarly good agreement between numerical results and the prediction in  Eq. \eqref{eq:Cassie_Baxter} is also obtained for other Young's contact angles.

\begin{figure*}
	\centering
	\includegraphics[width=\linewidth,keepaspectratio]{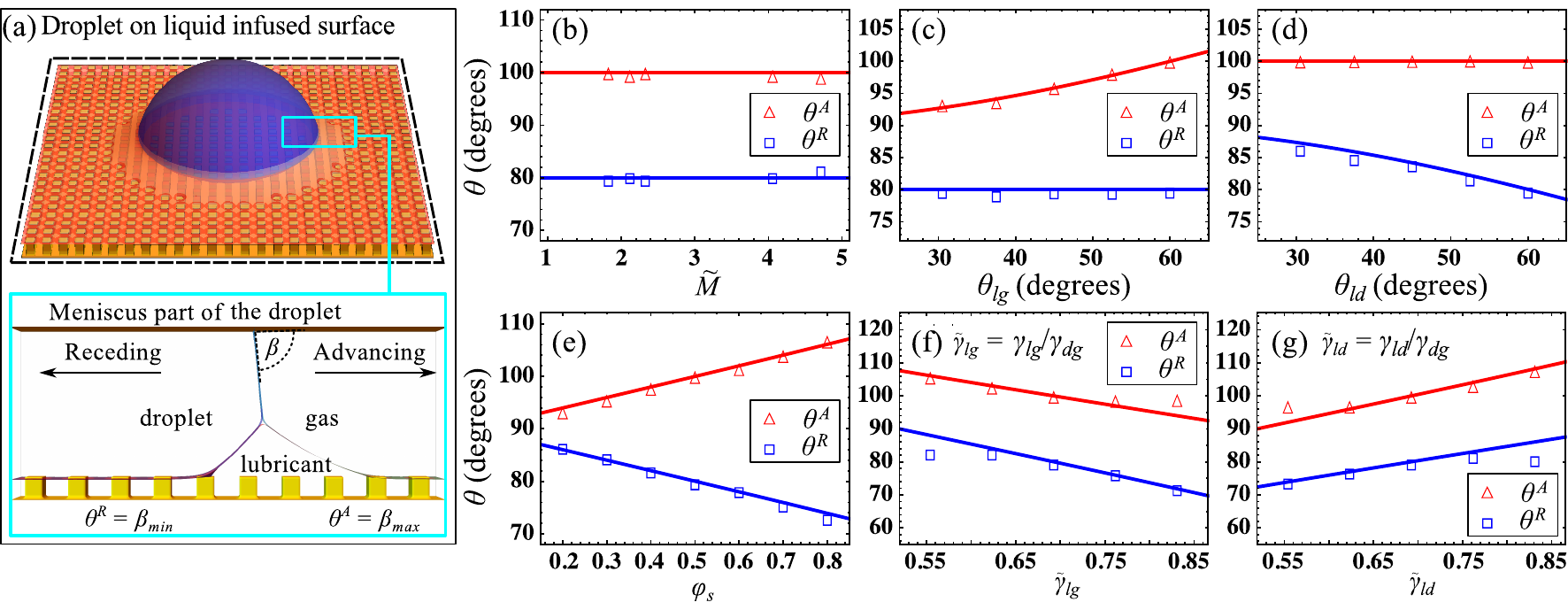}
	\caption{(a) Simulation setup for the advancing and the receding angles. To reduce computational costs, we focus on simulating the region around the lubricant meniscus. By varying $\beta$, we are able to investigate when the meniscus advances or recedes. The parameters studied are (b) the meniscus size, (c) the lubricant-gas wetting angle, (d) the lubricant-droplet wetting angle, (e) the solid fraction, and the ratios of (f) lubricant-gas and (g) lubricant-droplet surface tensions with the droplet-gas surface tension. In (b-g), the red and blue lines are theoretical predictions for $\theta^A$ and $\theta^R$ given in Eqs. \eqref{eq:receding} and \eqref{eq:advancing} respectively. The default values of the parameters are $\theta_{lg}=60^{\circ}$; $\theta_{ld}=60^{\circ}$; $\varphi_s=0.5$; and $\gamma_{lg}/\gamma_{dg} = \gamma_{ld}/\gamma_{dg} = 0.69$.}    
	\label{fig:fig3}
\end{figure*}

\subsubsection{The Advancing and the Receding Angles}
Following the previous subsection, the depinning angles for the advancing lubricant-gas and the receding lubricant-droplet interfaces are given by
\begin{align}
\cos\theta_{ld}^{depin}&=\varphi_s\cos\theta_{ld}+(1-\varphi_s),
\label{eq:ld_depining_angle}\\
\cos\theta_{lg}^{depin}&=\varphi_s\cos\theta_{lg}+(1-\varphi_s).
\label{eq:lg_depining_angle}
\end{align}
Substituting Eqs. \eqref{eq:ld_depining_angle} and \eqref{eq:lg_depining_angle}, as well as $[\theta_{lg}^{eff}]^R= [\theta_{ld}^{eff}]^A=0$, into Eq. \eqref{eq:fpinning2}, we obtain the full expression of the pinning force per unit length for a droplet on LIS:
\begin{align}
f_{LIS}=&\left[\gamma_{lg}-\gamma_{ld}\left(\varphi_s\cos\theta_{ld}+(1-\varphi_s)\right)\right]^R-\nonumber\\
&\left[\gamma_{lg}\left(\varphi_s\cos\theta_{lg}+(1-\varphi_s)\right)-\gamma_{ld}\right]^A, 
\label{eq:fpinning4}
\end{align}
{
	or alternatively,
	\begin{equation}
	f_{LIS}= \varphi_s \left[\gamma_{lg} (1-\cos\theta_{lg}) + \gamma_{ld} (1-\cos\theta_{ld})\right].
	\label{eq:fpinning4_2}
	\end{equation}
}
One important observation from Eq.~\eqref{eq:fpinning4} {or Eq.~\eqref{eq:fpinning4_2}} is that the magnitude of the pinning force does not actually depend on the droplet-gas surface tension, $\gamma_{dg}$, which distinguishes the case of pinning on LIS to pinning on other solid surfaces. {Additionally, Eq.~\eqref{eq:fpinning4_2} highlights that there is no pinning for the complete lubricant wetting case ($\theta_{lg}=\theta_{ld}=0^\circ$).} Nonetheless, to allow comparisons with other solid surfaces, it is useful to write Eq.~\eqref{eq:fpinning4} in the following form
\begin{align}
f_{LIS}=\gamma_{dg}\Delta\cos\theta,
\label{eq:fpinning5}
\end{align}
where $\Delta\cos\theta=\cos\theta^R-\cos\theta^A$, and the receding and the advancing contact angles are respectively defined as
\begin{align}
\cos\theta^R=& \frac{\gamma_{lg}}{\gamma_{dg}}-\frac{\gamma_{ld}}{\gamma_{dg}}(\varphi_s\cos\theta_{ld}+(1-\varphi_s)),\label{eq:receding}\\
\cos\theta^A=&\frac{\gamma_{lg}}{\gamma_{dg}}(\varphi_s\cos\theta_{lg}+(1-\varphi_s))-\frac{\gamma_{ld}}{\gamma_{dg}}.\label{eq:advancing}
\end{align}
These receding and the advancing contact angles are interpreted as the apparent contact angles at the front and rear of the lubricant meniscus as a liquid droplet depins on LIS (see Fig. \ref{fig:fig1}). Here, we also define CAH as $\Delta\theta=\theta^A-\theta^R$.

It is worth noting that, in this work, we have focussed on the case where the lubricant does not encapsulate the droplet. When the lubricant encapsulates the droplet, the effective droplet-gas surface tension becomes $\gamma_{dg}^{eff} = \gamma_{lg}+\gamma_{ld}$ \cite{McHale_2019}.

\subsection{The Effect of Changing Fluid and Solid Properties}
Eqs. \eqref{eq:receding} and \eqref{eq:advancing} suggest the advancing and receding angles are controlled by the surface tensions ($\gamma_{dg}, \gamma_{lg}, \gamma_{ld}$), the lubricant wetting angles ($\theta_{lg}, \theta_{ld}$), and the fraction of solid $\varphi_s$. In this subsection we will systematically test the validity and accuracy of Eqs. \eqref{eq:receding} and \eqref{eq:advancing} for predicting the advancing and receding angles.

To do this, rather than simulating the whole droplet (top panel of Fig. \ref{fig:fig3}(a)), we will focus on the region around the lubricant meniscus (lower panel of Fig. \ref{fig:fig3}(a)). In this simulation setup, the three fluid phases are present; and as in the setup in Fig. \ref{fig:fig2}, the movement of the meniscus is controlled by the contact angle at the top plate, $\beta$. The maximum angle $\beta_{max}$ for which the meniscus remains stable corresponds to the advancing angle $\theta^A$; while the minimum angle $\beta_{min}$ is the receding angle $\theta^R$.

\begin{figure*}[t]
	\centering
	\includegraphics[width=0.85\linewidth,keepaspectratio]{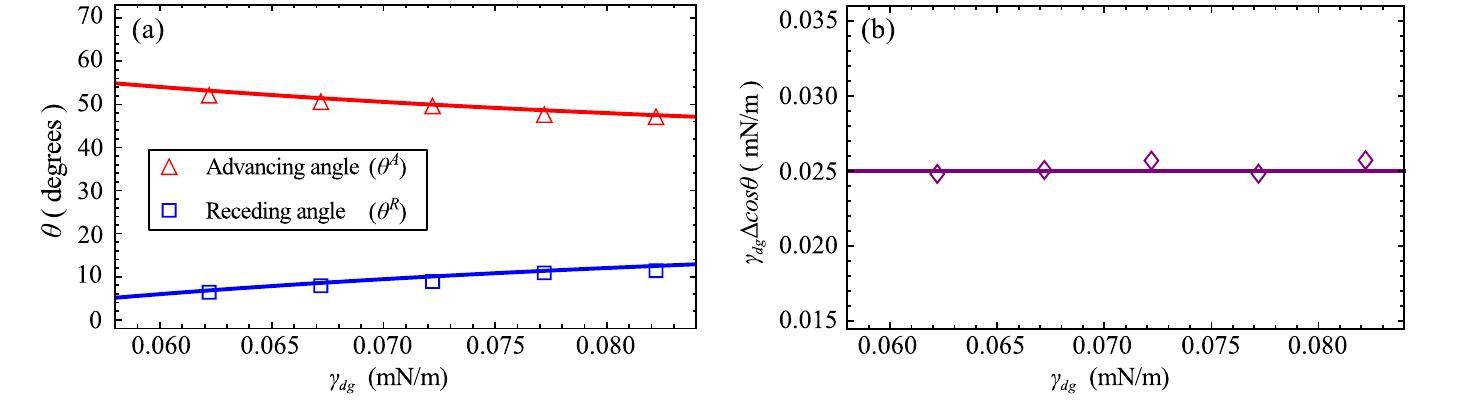}
	\caption{(a) The effect of $\gamma_{dg}$ on $\theta^A$ and $\theta^R$. The red and blue lines are the theoretical predictions for $\theta^A$ and $\theta^R$ as given in Eqs. \eqref{eq:receding} and \eqref{eq:advancing} respectively. (b) The pinning force per unit length for a droplet on LIS is independent of $\gamma_{dg}$.}
	\label{fig:fig5}
\end{figure*}

We first investigate the role of meniscus size on the advancing and receding angles. The meniscus size can be controlled by tuning the volume of the lubricant phase. Here, we parameterise the meniscus size $\tilde{M}$ by taking the ratio of the cross sectional area of the lubricant meniscus to the unit cell of the post. {The unit cell of the post is defined as the product between the centre-to-centre distance between neighbouring posts and the height of the posts}. Furthermore, we set the pressure in the droplet and gas phases to be equal, such that we are always in the vanishing meniscus regime \cite{Semprebon2016b} where the radius of the curvature of the lubricant meniscus is much smaller compared to the radius of curvature for the droplet-gas interface. From Fig. \ref{fig:fig3}(b) we can see that the advancing and the receding angles are independent of the meniscus size in this limit. 

The effect of the lubricant wetting angles, $\theta_{lg}$ and $\theta_{ld}$, are presented in panels (c) and (d) of Fig. \ref{fig:fig3}. In panel (c), we observe that $\theta_{lg}$ {\it only} affects the advancing angle but {\it not} the receding angle. This is because $\theta_{lg}$ controls the depinning angle of the lubricant-gas interface when the droplet is advancing. When the droplet is receding, the lubricant-gas interface moves by bridging the neighbouring post, which is independent from $\theta_{lg}$. Similarly, the bridging mechanism occurs for the lubricant-droplet interface during the advancing motion. As such, $\theta_{ld}$ does not affect the advancing angle, as shown in panel (d). In contrast, during the receding process, the lubricant-droplet interface moves by depinning from the post. Hence, the receding angle is affected by $\theta_{ld}$.

The influence of the solid fraction $\varphi_s$ is shown in panel (e) of Fig. \ref{fig:fig3}. {Here, we vary $\varphi_s=W_y/L_y$ by changing the post width in the direction perpendicular to the row of post ($W_y$).} It is intuitive to foresee that $\Delta\theta$ increases with $\varphi_s$. More specifically, this is because $\theta^A$ increases while $\theta^R$ decreases with $\varphi_s$. This finding is aligned with the experimental results in Ref. \cite{Smith2013}. In their work, although $\theta^A$ and $\theta^R$ were not measured directly, they showed that the pinning force that acts on a droplet on LIS can be reduced by employing surfaces with smaller $\varphi_s$ \cite{Smith2013}.

Next, the effect of the lubricant interfacial tensions is demonstrated in Fig. \ref{fig:fig3}(f) and (g). Interestingly, increasing $\gamma_{lg}$ decreases both the advancing and the receding angles, while for $\gamma_{ld}$, the effect is reversed. This is due to the fact that increasing $\gamma_{lg}$ generally makes a droplet on LIS to be more {\it hydrophilic}-like, while increasing $\gamma_{ld}$ makes it more {\it hydrophobic}-like, and thus the change of the contact angles follow accordingly \cite{Schellenberger2015,Sadullah2018}.

Finally, we have argued in Eq. \eqref{eq:fpinning4} {and Eq. \eqref{eq:fpinning4_2}} that the pinning force of a droplet on LIS does not depend on the droplet-gas interfacial tension $\gamma_{dg}$. Indeed, while the magnitudes of the advancing and receding angles are influenced by $\gamma_{dg}$, see Fig. \ref{fig:fig5}(a), the pinning force per unit length is constant regardless of $\gamma_{dg}$, as shown in Fig. \ref{fig:fig5}(b).

By studying each independent variable systematically, we have therefore demonstrated that Eqs. \eqref{eq:receding} and \eqref{eq:advancing} are an excellent model to describe the advancing and receding angles, as well as the contact angle hysteresis. All simulation results are in excellent agreement with this model.

\subsection{The Relationship between CAH, Sliding Angle and \\Pinning Force}
{One common practice to determine the CAH is to measure the advancing and the receding angles when a droplet is sliding as the substrate is inclined \cite{Eral2013}.} The droplet starts to slide when the external body force is larger than the pinning force that holds the droplet on the surface. The sliding angle $\alpha$ is related to the external body force via a simple relation
\begin{align}
F_{ext}=\rho V_{drop}g\sin\alpha.
\label{eq:f_ext}
\end{align}
Here, $\rho$ and $V_{drop}$ are the density and volume of the droplet, while $g$ is the gravitational acceleration. 

\begin{figure}
	\centering
	\includegraphics[width=\linewidth,keepaspectratio]{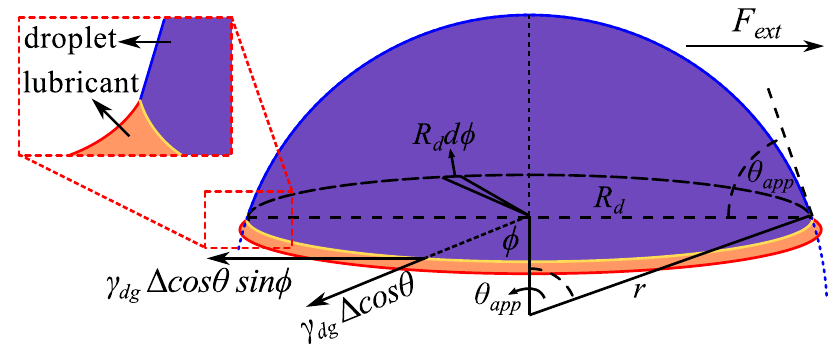}
	\caption{A sketch for the derivation of the pinning force of a droplet on LIS.}
	\label{fig:fig7}
\end{figure}

To obtain the total pinning force, we need to integrate the pinning force per unit length over the base perimeter of the droplet contact area with the solid. Consider the geometry illustrated in Fig. \ref{fig:fig7}, where again we focus on the vanishing meniscus limit. The pinning force per unit length is given by $\gamma_{dg}\Delta\cos\theta$; however, to balance the external force, we only need the vector component in the opposite direction of $F_{ext}$. Denoting $\phi$ as the azimuthal angle around the droplet, this corresponds to $\gamma_{dg}\Delta\cos\theta\sin\phi$. Assuming that the droplet base is circular, the total pinning force is then
\begin{equation}
F_{pin} =\int_{0}^{\pi}\gamma_{dg}\Delta\cos\theta\sin\phi R_dd\phi =2R_d\gamma_{dg}\Delta\cos\theta,
\label{eq:fpinning6}
\end{equation}
where $R_d$ is the droplet base radius. When $V_{drop}$ is known, $R_d$ is linked to the droplet apparent contact angle $\theta_{app}$ via
\begin{equation}
R_d =\left(\frac{(12/\pi)V_{drop}}{8-9\cos\theta_{app}+\cos(3\theta_{app})}\right)^{1/3}\sin\theta_{app},
\label{eq:base_radius}
\end{equation}
and the pinning force can be rewritten as
\begin{align}
F_{pin}=2\sin\theta_{app}\gamma_{dg}\Delta\cos\theta\left(\frac{(12/\pi)V_{drop}}{8-9\cos\theta_{app}+\cos(3\theta_{app})}\right)^{1/3}.
\label{eq:fpinning7}
\end{align}

We can now balance the external body force Eq. \eqref{eq:f_ext} with the pinning force Eq. \eqref{eq:fpinning6} to obtain the theoretical prediction of the sliding angle $\alpha$, which is given by
\begin{align}
\alpha=\sin^{-1}\left(\frac{2\gamma_{dg}R_d\Delta\cos\theta}{\rho V_{drop}g}\right).
\label{eq:sliding_angle}
\end{align}
Using Eq. \eqref{eq:sliding_angle}, we can compare our theoretical prediction against available experimental results. For this purpose, we use the experimental data reported in Ref. \cite{Smith2013} for water droplet on BMIm (an ionic liquid) infused surface. The comparisons are given in Table \ref{tab:tab1}. The predicted sliding angles are consistent with the experimental values $\alpha_{e}$.

\begin{table}[h]
	\small
	\caption{Comparisons between the experimental data from Ref. \cite{Smith2013} and the theoretical predictions using Eq. \eqref{eq:sliding_angle} for the sliding angles of droplets on LIS.} 
	\label{tab:tab1}
	\begin{tabular*}{0.48\textwidth}{@{\extracolsep{\fill}}ccccc}
		\hline
		$\varphi_s$ & $\Delta\theta~(^\circ)$ & $\alpha~(^\circ)$ & $\alpha_{e}~(^\circ)$ & $|\alpha - \alpha_{e}|~(^\circ)$\\
		\hline
		0.25    &    8    &    28 & 30 & 2\\
		0.33    &    11    &    37 & 45 & 8\\
		0.44    &    14    &    53 & 60 & 7\\
		\hline
	\end{tabular*}
\end{table}

From Table \ref{tab:tab1} we find that, on LIS, a relatively low CAH ($\Delta\theta$) can still lead to a significant critical sliding angle $\alpha$. This is different compared to superhydrophobic surfaces where $\alpha$ usually has the same magnitude as $\Delta\theta$. It also suggests we should be cautious when using $\Delta\theta$ to characterise the mobility (and more generally, liquid repellency) of a liquid droplet on LIS.

{To explain why a droplet on LIS may suffer from a large pinning force, let us now consider two key aspects in which the contact angle of a droplet can affect the pinning force in Eq. \eqref{eq:fpinning6}, namely through the droplet base perimeter $R$ and the difference in the cosine of the contact angle $\Delta \cos\theta$. First, for LIS, the apparent contact angle $\theta_{app}$ is relatively low such that the droplet base perimeter is large, in direct contrast to the large contact angle and small base perimeter for drops on classical superhydrophobic surfaces. This large droplet base perimeter can potentially magnify the pinning force, since $F_{pin} \propto R$. Second, the $\Delta\cos\theta$ term has an implicit dependence on the contact angle. Even for the same value of $\Delta\theta$, $\Delta\cos\theta$ is greater when $\theta_{app}\approx90^\circ$ than when $\theta_{app}\approx180^\circ$ or $\theta_{app}\approx0^\circ$. Therefore, droplets on LIS are prone to large pinning forces when $\Delta\theta$ is large since most LIS systems reported in the literature have $\theta_{app}\approx90^\circ$.}

It is useful to express the pinning force in a non-dimensionalised form, given by
\begin{align}
\tilde{F}_{pin}&=\frac{F_{pin}}{\gamma_{dg}\sqrt[3]{V_{drop}}},\\
\tilde{F}_{pin}&\simeq\sqrt[3]{\frac{(12/\pi)(2\sin^2\theta_{app}\Delta\theta)^3}{8-9\cos\theta_{app}+\cos(3\theta_{app})}}
\label{eq:fpinning8}
\end{align}
for small $\Delta \theta$. This non-dimensionalised form of the pinning force depends only on $\Delta\theta$ and $\theta_{app}$, which respectively represent the CAH and the shape of the droplet. {It is worth noting that Eq.~\eqref{eq:fpinning8} is generally valid for any surface, not just for LIS.}

\begin{figure}
	\centering
	\includegraphics[width=\linewidth,keepaspectratio]{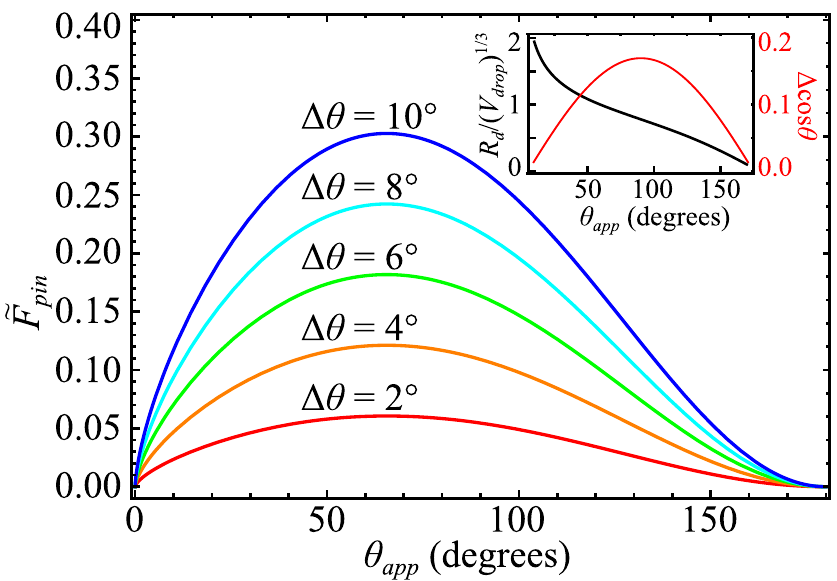}
	\caption{Visualisation of the effects of CAH and the droplet shape on the pinning force. The inset shows the non-dimensionalised droplet base radius $R_d/\sqrt[3]{V_{drop}}$ and $\Delta\cos\theta$ against $\theta_{app}$.}
	\label{fig:fig8}
\end{figure}

The effects of $\Delta\theta$ and $\theta_{app}$ on the pinning force are visualised in Fig. \ref{fig:fig8}. Interestingly, we find that the pinning force reaches its maximum at $\theta_{app}=65.5^\circ$, regardless of $\Delta\theta$. Therefore, for LIS, it is advisable to avoid the droplet-lubricant combinations which result in $\theta_{app}\approx65.5^\circ$. The $\Delta\cos\theta$ itself reaches its maximum at $\theta_{app}=90^\circ$ for any given value of $\Delta\theta$, as shown as the red plot in the inset of Fig. \ref{fig:fig8}. This is an indication that $\Delta\cos\theta$ is not the only factor that controls the pinning force. The shift in the maximum of $\tilde{F}_{pin}$ to the lower $\theta_{app}$ is due to the contribution from the droplet base perimeter. As shown in the inset (black plot), the non-dimensionalised droplet base radius $R_d/(V_{drop})^{1/3}$ is larger for smaller $\theta_{app}$.

{Consistent with our theory,} Fig. \ref{fig:fig8} also rationalises why pinning force is small for superhydrophobic surfaces. This is because both $\Delta\cos\theta$ and $R_d/(V_{drop})^{1/3}$ go to zero as $\theta \rightarrow 180^\circ$.

\section{Conclusions}
In this work, we have considered the contact angle hysteresis and pinning force of a droplet on LIS {in the limit of the vanishing meniscus case}. We have derived the expressions for the advancing and receding angles as well as the pinning force using a force balance argument, including how they depend on the liquid interfacial tensions, the lubricant wetting angles, and the solid fraction. Each dependency was systematically tested and verified using numerical simulations based on a diffuse interface approach. We also found that the pinning force does not depend on the droplet-gas interfacial tension.

We have also derived an analytical expression for the critical droplet sliding angle, and the predictions from our theory are consistent with experimental data reported by Smith \textit{et al.} \cite{Smith2013}. Furthermore, using this theory, we assess why liquid droplets on LIS suffer from larger pinning forces compared to superhydrophobic surfaces, even for the same $\Delta\theta$. We conclude this is due to two factors: both the droplet base perimeter and the magnitude of $\Delta\cos\theta$ are typically larger in LIS due to the lower (apparent) contact angle.

This study helps us to carefully design LIS by providing insights into how each relevant parameter influences the pinning force. Although the example shown here is for the {rectangular} posts, similar derivations of the pinning force, as well as the advancing and the receding angles, can also be done for different surface geometries by following the same approach. Interestingly, the derivations rely on the depinning mechanism of each lubricant interface, which is just a binary fluid case. This shows an example where the complexity of ternary fluids systems can be broken down into their constituting binary fluids problems. It would be therefore interesting for future research to test our theory for more complex geometries{, in particular for regimes where the Cassie-Baxter approximation is known to break down for contact angle hysteresis on superhydrophobic surfaces {\cite{Joanny1984,Reyssat2009,Raj2012}.}} Furthermore, we hope our theory will motivate systematic experimental verifications, harnessing recent advances in surface fabrication techniques for LIS.

\section{Acknowledgements}
MSS is supported by an LPDP (Lembaga Pengelola Dana Pendidikan) scholarship from the Indonesian Government. JRP and HK acknowledge funding from EPSRC (grant EP/P007139/1) and Procter and Gamble. We thank to Dr Ciro Semprebon, Dr Yonas Gizaw and Dr Dan Daniel for the fruitful discussions.


\bibliographystyle{rsc} 
\bibliography{template.article}

\providecommand*{\mcitethebibliography}{\thebibliography}
\csname @ifundefined\endcsname{endmcitethebibliography}
{\let\endmcitethebibliography\endthebibliography}{}
\begin{mcitethebibliography}{38}
\providecommand*{\natexlab}[1]{#1}
\providecommand*{\mciteSetBstSublistMode}[1]{}
\providecommand*{\mciteSetBstMaxWidthForm}[2]{}
\providecommand*{\mciteBstWouldAddEndPuncttrue}
  {\def\EndOfBibitem{\unskip.}}
\providecommand*{\mciteBstWouldAddEndPunctfalse}
  {\let\EndOfBibitem\relax}
\providecommand*{\mciteSetBstMidEndSepPunct}[3]{}
\providecommand*{\mciteSetBstSublistLabelBeginEnd}[3]{}
\providecommand*{\EndOfBibitem}{}
\mciteSetBstSublistMode{f}
\mciteSetBstMaxWidthForm{subitem}
{(\emph{\alph{mcitesubitemcount}})}
\mciteSetBstSublistLabelBeginEnd{\mcitemaxwidthsubitemform\space}
{\relax}{\relax}

\bibitem[Qu{\'{e}}r{\'{e}}(2005)]{Quere2005}
D.~Qu{\'{e}}r{\'{e}}, \emph{Rep. Prog. Phys.}, 2005, \textbf{68},
  2495--2532\relax
\mciteBstWouldAddEndPuncttrue
\mciteSetBstMidEndSepPunct{\mcitedefaultmidpunct}
{\mcitedefaultendpunct}{\mcitedefaultseppunct}\relax
\EndOfBibitem
\bibitem[Wong \emph{et~al.}(2011)Wong, Kang, Tang, Smythe, Hatton, Grinthal,
  and Aizenberg]{Wong2011}
T.-S. Wong, S.~H. Kang, S.~K.~Y. Tang, E.~J. Smythe, B.~D. Hatton, A.~Grinthal
  and J.~Aizenberg, \emph{Nature}, 2011, \textbf{477}, 443--7\relax
\mciteBstWouldAddEndPuncttrue
\mciteSetBstMidEndSepPunct{\mcitedefaultmidpunct}
{\mcitedefaultendpunct}{\mcitedefaultseppunct}\relax
\EndOfBibitem
\bibitem[Smith \emph{et~al.}(2013)Smith, Dhiman, Anand, Reza-Garduno, Cohen,
  McKinley, and Varanasi]{Smith2013}
J.~D. Smith, R.~Dhiman, S.~Anand, E.~Reza-Garduno, R.~E. Cohen, G.~H. McKinley
  and K.~K. Varanasi, \emph{Soft Matter}, 2013, \textbf{9}, 1772--1780\relax
\mciteBstWouldAddEndPuncttrue
\mciteSetBstMidEndSepPunct{\mcitedefaultmidpunct}
{\mcitedefaultendpunct}{\mcitedefaultseppunct}\relax
\EndOfBibitem
\bibitem[Sotiri \emph{et~al.}(2016)Sotiri, Overton, Waterhouse, and
  Howell]{Sotiri2016}
I.~Sotiri, J.~C. Overton, A.~Waterhouse and C.~Howell, \emph{Exp. Biol. Med.},
  2016, \textbf{241}, 909--918\relax
\mciteBstWouldAddEndPuncttrue
\mciteSetBstMidEndSepPunct{\mcitedefaultmidpunct}
{\mcitedefaultendpunct}{\mcitedefaultseppunct}\relax
\EndOfBibitem
\bibitem[Ware \emph{et~al.}(2018)Ware, Smith-Palmer, Peppou-Chapman, Scarratt,
  Humphries, Balzer, and Neto]{Ware2018}
C.~S. Ware, T.~Smith-Palmer, S.~Peppou-Chapman, L.~R. Scarratt, E.~M.
  Humphries, D.~Balzer and C.~Neto, \emph{ACS Appl. Mater. Interfaces}, 2018,
  \textbf{10}, 4173--4182\relax
\mciteBstWouldAddEndPuncttrue
\mciteSetBstMidEndSepPunct{\mcitedefaultmidpunct}
{\mcitedefaultendpunct}{\mcitedefaultseppunct}\relax
\EndOfBibitem
\bibitem[Brown and Bhushan(2017)]{Brown2017}
P.~S. Brown and B.~Bhushan, \emph{J. Colloid Interface Sci.}, 2017,
  \textbf{487}, 437--443\relax
\mciteBstWouldAddEndPuncttrue
\mciteSetBstMidEndSepPunct{\mcitedefaultmidpunct}
{\mcitedefaultendpunct}{\mcitedefaultseppunct}\relax
\EndOfBibitem
\bibitem[Geng and Cho(2019)]{Geng2019}
H.~Geng and S.~K. Cho, \emph{Lab Chip}, 2019, \textbf{19}, 2275--2283\relax
\mciteBstWouldAddEndPuncttrue
\mciteSetBstMidEndSepPunct{\mcitedefaultmidpunct}
{\mcitedefaultendpunct}{\mcitedefaultseppunct}\relax
\EndOfBibitem
\bibitem[Wei \emph{et~al.}(2014)Wei, Schlaich, Pr{\'{e}}vost, Schulz,
  B{\"{o}}ttcher, Gradzielski, Qi, Haag, and Schalley]{Wei2014}
Q.~Wei, C.~Schlaich, S.~Pr{\'{e}}vost, A.~Schulz, C.~B{\"{o}}ttcher,
  M.~Gradzielski, Z.~Qi, R.~Haag and C.~A. Schalley, \emph{Adv. Mater.}, 2014,
  \textbf{26}, 7358--7364\relax
\mciteBstWouldAddEndPuncttrue
\mciteSetBstMidEndSepPunct{\mcitedefaultmidpunct}
{\mcitedefaultendpunct}{\mcitedefaultseppunct}\relax
\EndOfBibitem
\bibitem[Galvan \emph{et~al.}(2018)Galvan, Phillips, Haumann, Wasserscheid,
  Zarraga, and Vogel]{Galvan2018}
Y.~Galvan, K.~R. Phillips, M.~Haumann, P.~Wasserscheid, R.~Zarraga and
  N.~Vogel, \emph{Langmuir}, 2018, \textbf{34}, 6894--6902\relax
\mciteBstWouldAddEndPuncttrue
\mciteSetBstMidEndSepPunct{\mcitedefaultmidpunct}
{\mcitedefaultendpunct}{\mcitedefaultseppunct}\relax
\EndOfBibitem
\bibitem[Howell \emph{et~al.}(2015)Howell, Vu, Johnson, Hou, Ahanotu,
  Alvarenga, Leslie, Uzun, Waterhouse, Kim, Super, Aizenberg, Ingber, and
  Aizenberg]{Howell2015}
C.~Howell, T.~L. Vu, C.~P. Johnson, X.~Hou, O.~Ahanotu, J.~Alvarenga, D.~C.
  Leslie, O.~Uzun, A.~Waterhouse, P.~Kim, M.~Super, M.~Aizenberg, D.~E. Ingber
  and J.~Aizenberg, \emph{Chem. Mater.}, 2015, \textbf{27}, 1792--1800\relax
\mciteBstWouldAddEndPuncttrue
\mciteSetBstMidEndSepPunct{\mcitedefaultmidpunct}
{\mcitedefaultendpunct}{\mcitedefaultseppunct}\relax
\EndOfBibitem
\bibitem[Liu \emph{et~al.}(2016)Liu, Wexler, Sch{\"{o}}necker, and
  Stone]{Liu2016}
Y.~Liu, J.~S. Wexler, C.~Sch{\"{o}}necker and H.~A. Stone, \emph{Phys. Rev.
  Fluids}, 2016, \textbf{1}, 074003\relax
\mciteBstWouldAddEndPuncttrue
\mciteSetBstMidEndSepPunct{\mcitedefaultmidpunct}
{\mcitedefaultendpunct}{\mcitedefaultseppunct}\relax
\EndOfBibitem
\bibitem[Sett \emph{et~al.}(2017)Sett, Yan, Barac, Bolton, and
  Miljkovic]{Sett2017}
S.~Sett, X.~Yan, G.~Barac, L.~W. Bolton and N.~Miljkovic, \emph{ACS Appl.
  Mater. Interfaces}, 2017, \textbf{9}, 36400--36408\relax
\mciteBstWouldAddEndPuncttrue
\mciteSetBstMidEndSepPunct{\mcitedefaultmidpunct}
{\mcitedefaultendpunct}{\mcitedefaultseppunct}\relax
\EndOfBibitem
\bibitem[Peppou-Chapman and Neto(2018)]{Peppou-Chapman2018}
S.~Peppou-Chapman and C.~Neto, \emph{ACS Appl. Mater. Interfaces}, 2018,
  \textbf{10}, 33669--33677\relax
\mciteBstWouldAddEndPuncttrue
\mciteSetBstMidEndSepPunct{\mcitedefaultmidpunct}
{\mcitedefaultendpunct}{\mcitedefaultseppunct}\relax
\EndOfBibitem
\bibitem[Kreder \emph{et~al.}(2018)Kreder, Daniel, Tetreault, Cao, Lemaire,
  Timonen, and Aizenberg]{Kreder2018}
M.~J. Kreder, D.~Daniel, A.~Tetreault, Z.~Cao, B.~Lemaire, J.~V. Timonen and
  J.~Aizenberg, \emph{Phys. Rev. X}, 2018, \textbf{8}, 031053\relax
\mciteBstWouldAddEndPuncttrue
\mciteSetBstMidEndSepPunct{\mcitedefaultmidpunct}
{\mcitedefaultendpunct}{\mcitedefaultseppunct}\relax
\EndOfBibitem
\bibitem[Lou \emph{et~al.}(2020)Lou, Huang, Yang, Zhu, Heng, and Xia]{Lou2020}
X.~Lou, Y.~Huang, X.~Yang, H.~Zhu, L.~Heng and F.~Xia, \emph{Adv. Funct.
  Mater.}, 2020, \textbf{30}, 1901130\relax
\mciteBstWouldAddEndPuncttrue
\mciteSetBstMidEndSepPunct{\mcitedefaultmidpunct}
{\mcitedefaultendpunct}{\mcitedefaultseppunct}\relax
\EndOfBibitem
\bibitem[Wang \emph{et~al.}(2017)Wang, Qian, Lai, Wang, Ma, Guo, Zhu, Fei, and
  Xin]{Wang2017}
Y.~Wang, B.~Qian, C.~Lai, X.~Wang, K.~Ma, Y.~Guo, X.~Zhu, B.~Fei and J.~H. Xin,
  \emph{ACS Appl. Mater. Interfaces}, 2017, \textbf{9}, 24428--24432\relax
\mciteBstWouldAddEndPuncttrue
\mciteSetBstMidEndSepPunct{\mcitedefaultmidpunct}
{\mcitedefaultendpunct}{\mcitedefaultseppunct}\relax
\EndOfBibitem
\bibitem[Sadullah \emph{et~al.}(2020)Sadullah, Launay, Parle, Ledesma-Aguilar,
  Gizaw, McHale, Wells, and Kusumaatmaja]{Sadullah2020}
M.~S. Sadullah, G.~Launay, J.~Parle, R.~Ledesma-Aguilar, Y.~Gizaw, G.~McHale,
  G.~G. Wells and H.~Kusumaatmaja, \emph{arXiv}, 2020,  2004.10408\relax
\mciteBstWouldAddEndPuncttrue
\mciteSetBstMidEndSepPunct{\mcitedefaultmidpunct}
{\mcitedefaultendpunct}{\mcitedefaultseppunct}\relax
\EndOfBibitem
\bibitem[Wang \emph{et~al.}(2018)Wang, Timonen, Carlson, Drotlef, Zhang, Kolle,
  Grinthal, Wong, Hatton, Kang, Kennedy, Chi, Blough, Sitti, Mahadevan, and
  Aizenberg]{Wang2018}
W.~Wang, J.~V.~I. Timonen, A.~Carlson, D.-M. Drotlef, C.~T. Zhang, S.~Kolle,
  A.~Grinthal, T.-S. Wong, B.~Hatton, S.~H. Kang, S.~Kennedy, J.~Chi, R.~T.
  Blough, M.~Sitti, L.~Mahadevan and J.~Aizenberg, \emph{Nature}, 2018,
  \textbf{559}, 77--82\relax
\mciteBstWouldAddEndPuncttrue
\mciteSetBstMidEndSepPunct{\mcitedefaultmidpunct}
{\mcitedefaultendpunct}{\mcitedefaultseppunct}\relax
\EndOfBibitem
\bibitem[Semprebon \emph{et~al.}(2016)Semprebon, Kr{\"{u}}ger, and
  Kusumaatmaja]{Semprebon2016a}
C.~Semprebon, T.~Kr{\"{u}}ger and H.~Kusumaatmaja, \emph{Phys. Rev. E}, 2016,
  \textbf{93}, 033305\relax
\mciteBstWouldAddEndPuncttrue
\mciteSetBstMidEndSepPunct{\mcitedefaultmidpunct}
{\mcitedefaultendpunct}{\mcitedefaultseppunct}\relax
\EndOfBibitem
\bibitem[K.~Connington(2013)]{Connington2013}
T.~L. K.~Connington, \emph{J. Comp. Phys.}, 2013, \textbf{250}, 601--615\relax
\mciteBstWouldAddEndPuncttrue
\mciteSetBstMidEndSepPunct{\mcitedefaultmidpunct}
{\mcitedefaultendpunct}{\mcitedefaultseppunct}\relax
\EndOfBibitem
\bibitem[Girifalco and Good(1957)]{Girifalco1957}
L.~Girifalco and R.~Good, \emph{J. Phys. Chem.}, 1957, \textbf{61},
  904--909\relax
\mciteBstWouldAddEndPuncttrue
\mciteSetBstMidEndSepPunct{\mcitedefaultmidpunct}
{\mcitedefaultendpunct}{\mcitedefaultseppunct}\relax
\EndOfBibitem
\bibitem[Liu and Nocedal(1989)]{Liu1989}
D.~C. Liu and J.~Nocedal, \emph{Math. Program.}, 1989, \textbf{45},
  503--528\relax
\mciteBstWouldAddEndPuncttrue
\mciteSetBstMidEndSepPunct{\mcitedefaultmidpunct}
{\mcitedefaultendpunct}{\mcitedefaultseppunct}\relax
\EndOfBibitem
\bibitem[Kusumaatmaja(2015)]{Kusumaatmaja2015}
H.~Kusumaatmaja, \emph{J. Chem. Phys.}, 2015, \textbf{142}, 124112\relax
\mciteBstWouldAddEndPuncttrue
\mciteSetBstMidEndSepPunct{\mcitedefaultmidpunct}
{\mcitedefaultendpunct}{\mcitedefaultseppunct}\relax
\EndOfBibitem
\bibitem[Panter \emph{et~al.}(2019)Panter, Gizaw, and Kusumaatmaja]{Panter2019}
J.~R. Panter, Y.~Gizaw and H.~Kusumaatmaja, \emph{Sci. Adv.}, 2019, \textbf{5},
  eaav7328\relax
\mciteBstWouldAddEndPuncttrue
\mciteSetBstMidEndSepPunct{\mcitedefaultmidpunct}
{\mcitedefaultendpunct}{\mcitedefaultseppunct}\relax
\EndOfBibitem
\bibitem[Qu{\'{e}}r{\'{e}}(2008)]{Quere2008}
D.~Qu{\'{e}}r{\'{e}}, \emph{Annu. Rev. Mater. Res.}, 2008, \textbf{38},
  71--99\relax
\mciteBstWouldAddEndPuncttrue
\mciteSetBstMidEndSepPunct{\mcitedefaultmidpunct}
{\mcitedefaultendpunct}{\mcitedefaultseppunct}\relax
\EndOfBibitem
\bibitem[Semprebon \emph{et~al.}(2017)Semprebon, McHale, and
  Kusumaatmaja]{Semprebon2016b}
C.~Semprebon, G.~McHale and H.~Kusumaatmaja, \emph{Soft Matter}, 2017,
  \textbf{13}, 101--110\relax
\mciteBstWouldAddEndPuncttrue
\mciteSetBstMidEndSepPunct{\mcitedefaultmidpunct}
{\mcitedefaultendpunct}{\mcitedefaultseppunct}\relax
\EndOfBibitem
\bibitem[Cassie and Baxter(1944)]{CassieBaxter1944}
A.~Cassie and S.~Baxter, \emph{Trans. Faraday Soc.}, 1944, \textbf{40},
  546\relax
\mciteBstWouldAddEndPuncttrue
\mciteSetBstMidEndSepPunct{\mcitedefaultmidpunct}
{\mcitedefaultendpunct}{\mcitedefaultseppunct}\relax
\EndOfBibitem
\bibitem[Keiser \emph{et~al.}(2017)Keiser, Keiser, Clanet, and
  Qu{\'{e}}r{\'{e}}]{Keiser2017}
A.~Keiser, L.~Keiser, C.~Clanet and D.~Qu{\'{e}}r{\'{e}}, \emph{Soft Matter},
  2017, \textbf{13}, 6981--6987\relax
\mciteBstWouldAddEndPuncttrue
\mciteSetBstMidEndSepPunct{\mcitedefaultmidpunct}
{\mcitedefaultendpunct}{\mcitedefaultseppunct}\relax
\EndOfBibitem
\bibitem[Mognetti and Yeomans(2010)]{Mognetti2010}
B.~M. Mognetti and J.~M. Yeomans, \emph{Langmuir}, 2010, \textbf{26},
  18162--18168\relax
\mciteBstWouldAddEndPuncttrue
\mciteSetBstMidEndSepPunct{\mcitedefaultmidpunct}
{\mcitedefaultendpunct}{\mcitedefaultseppunct}\relax
\EndOfBibitem
\bibitem[Schellenberger \emph{et~al.}(2016)Schellenberger, Encinas, Vollmer,
  and Butt]{Schellenberger2016}
F.~Schellenberger, N.~Encinas, D.~Vollmer and H.-J. Butt, \emph{Phys. Rev.
  Lett.}, 2016, \textbf{116}, 096101\relax
\mciteBstWouldAddEndPuncttrue
\mciteSetBstMidEndSepPunct{\mcitedefaultmidpunct}
{\mcitedefaultendpunct}{\mcitedefaultseppunct}\relax
\EndOfBibitem
\bibitem[Priest \emph{et~al.}(2007)Priest, Sedev, and Ralston]{Priest2007}
C.~Priest, R.~Sedev and J.~Ralston, \emph{Phys. Rev. Lett.}, 2007, \textbf{99},
  026103\relax
\mciteBstWouldAddEndPuncttrue
\mciteSetBstMidEndSepPunct{\mcitedefaultmidpunct}
{\mcitedefaultendpunct}{\mcitedefaultseppunct}\relax
\EndOfBibitem
\bibitem[McHale \emph{et~al.}(2019)McHale, Orme, Wells, and
  Ledesma-Aguilar]{McHale_2019}
G.~McHale, B.~V. Orme, G.~G. Wells and R.~Ledesma-Aguilar, \emph{Langmuir},
  2019, \textbf{35}, 4197--4204\relax
\mciteBstWouldAddEndPuncttrue
\mciteSetBstMidEndSepPunct{\mcitedefaultmidpunct}
{\mcitedefaultendpunct}{\mcitedefaultseppunct}\relax
\EndOfBibitem
\bibitem[Schellenberger \emph{et~al.}(2015)Schellenberger, Xie, Encinas, Hardy,
  Klapper, Papadopoulos, Butt, and Vollmer]{Schellenberger2015}
F.~Schellenberger, J.~Xie, N.~Encinas, A.~Hardy, M.~Klapper, P.~Papadopoulos,
  H.-J. Butt and D.~Vollmer, \emph{Soft Matter}, 2015, \textbf{11},
  7617--7626\relax
\mciteBstWouldAddEndPuncttrue
\mciteSetBstMidEndSepPunct{\mcitedefaultmidpunct}
{\mcitedefaultendpunct}{\mcitedefaultseppunct}\relax
\EndOfBibitem
\bibitem[Sadullah \emph{et~al.}(2018)Sadullah, Semprebon, and
  Kusumaatmaja]{Sadullah2018}
M.~S. Sadullah, C.~Semprebon and H.~Kusumaatmaja, \emph{Langmuir}, 2018,
  \textbf{34}, 8112--8118\relax
\mciteBstWouldAddEndPuncttrue
\mciteSetBstMidEndSepPunct{\mcitedefaultmidpunct}
{\mcitedefaultendpunct}{\mcitedefaultseppunct}\relax
\EndOfBibitem
\bibitem[Eral \emph{et~al.}(2013)Eral, 't~Mannetje, and Oh]{Eral2013}
H.~B. Eral, D.~'t~Mannetje and J.~M. Oh, \emph{Colloid Polym. Sci.}, 2013,
  \textbf{291}, 247--260\relax
\mciteBstWouldAddEndPuncttrue
\mciteSetBstMidEndSepPunct{\mcitedefaultmidpunct}
{\mcitedefaultendpunct}{\mcitedefaultseppunct}\relax
\EndOfBibitem
\bibitem[Joanny and de~Gennes(1984)]{Joanny1984}
J.~F. Joanny and P.~G. de~Gennes, \emph{J. Chem. Phys}, 1984, \textbf{81},
  552--562\relax
\mciteBstWouldAddEndPuncttrue
\mciteSetBstMidEndSepPunct{\mcitedefaultmidpunct}
{\mcitedefaultendpunct}{\mcitedefaultseppunct}\relax
\EndOfBibitem
\bibitem[Reyssat and Quéré(2009)]{Reyssat2009}
M.~Reyssat and D.~Quéré, \emph{J. Phys. Chem. B}, 2009, \textbf{113},
  3906--3909\relax
\mciteBstWouldAddEndPuncttrue
\mciteSetBstMidEndSepPunct{\mcitedefaultmidpunct}
{\mcitedefaultendpunct}{\mcitedefaultseppunct}\relax
\EndOfBibitem
\bibitem[Raj \emph{et~al.}(2012)Raj, Enright, Zhu, Adera, and Wang]{Raj2012}
R.~Raj, R.~Enright, Y.~Zhu, S.~Adera and E.~N. Wang, \emph{Langmuir}, 2012,
  \textbf{28}, 15777--15788\relax
\mciteBstWouldAddEndPuncttrue
\mciteSetBstMidEndSepPunct{\mcitedefaultmidpunct}
{\mcitedefaultendpunct}{\mcitedefaultseppunct}\relax
\EndOfBibitem
\end{mcitethebibliography}


\end{document}